\documentclass[aps,prx,showpacs,amsmath,amssymb,amsfonts,lengthcheck,twocolumn,longbibliography,superscriptaddress, nofootinbib]{revtex4-2}

\usepackage{changes} 
\usepackage{cancel}
\usepackage{enumitem}
\usepackage{graphicx}
\usepackage{subfigure}
\usepackage{amsthm}
\usepackage{amsmath}
\usepackage{verbatim}
\usepackage{dcolumn}
\usepackage{bm}
\usepackage{color}
\usepackage[colorlinks=true,citecolor=blue,linkcolor=blue,urlcolor=blue]{hyperref}%
\usepackage{xcolor}
\usepackage{dsfont}
\usepackage{longtable}
\usepackage{tabularx}
\usepackage[ruled,vlined]{algorithm2e}
\allowdisplaybreaks
\newcommand{\bra}[1]{\left\langle #1\right|}
\newcommand{\ket}[1]{\left|#1\right\rangle}

\newcommand{\tr}[1]{\mathrm{tr}\left\{#1\right\}}

\newcommand{\blah}{blah\\blah\\blah\\blah\\blah}

\begin{document}

\title{Adaptive Dissipative State Preparation through Reinforcement Learning}
\date{\today}
\author{Nathan M. Myers}
\email{nathan.myers@pnnl.gov}
\affiliation{Physical and Computational Sciences, Pacific Northwest National Laboratory, Richland, WA 99354, USA}
\author{Chenxu Liu}
\affiliation {Physical and Computational Sciences, Pacific Northwest National Laboratory, Richland, WA 99354, USA}
\author{Yulong Dong}
\affiliation {Department of Electrical Engineering and Computer Science, University of Michigan, Ann Arbor, MI 48109, USA}
\author{Nicholas P. Bauman}
\affiliation{Physical and Computational Sciences, Pacific Northwest National Laboratory, Richland, WA 99354, USA}
\author{Karol Kowalski}
\affiliation {Physical and Computational Sciences, Pacific Northwest National Laboratory, Richland, WA 99354, USA}

\begin{abstract}
Dissipative algorithms approach the problem of ground state preparation by mimicking the natural thermalization of a quantum system in contact with a large, low-temperature thermal environment. The environment can be efficiently simulated by a single ancilla qubit with a variable energy gap that is repeatedly coupled to the system qubits to generate a dissipative channel, and then reset after each interaction. Here we present an adaptive implementation of the dissipative algorithm, RL-Adapt, that uses single-shot reinforcement learning to optimize the selection of ancilla frequency and system-bath interaction operator to maximize energy dissipation without relying on a priori knowledge of the system spectrum. The adaptive implementation results in significantly reduced convergence times and can successfully find the ground state even for non-ideal operator pools that fail to converge using non-adaptive, uniform random operator and ancilla frequency selection.      
\end{abstract}

\maketitle

\section{Introduction} 
\label{sec:1}

Simulating many-body quantum systems is one of the most promising applications for near-term quantum computers~\cite{ChildsPNAS2018, Alexeev2025JCTC}. The primary goal of many Hamiltonian simulation algorithms is to determine the spectrum, and in particular the ground state, of a many-body quantum system. Numerous methods have been proposed and implemented, including quantum phase estimation~\cite{Abrams1999PRL, AspuruGuzik2005Sci, Berry2007CMP, Ni2023Quantum, Ding2023PRXQ}, variational quantum eigensolver (VQE)~\cite{Peruzzo2014Nat, McClean2016NJP, Kandala2017Nat, Tilly2022PR} and its derivatives such as ADAPT-VQE~\cite{Grimsley2019Nat, Tang2021PRXQ, Tang2025PRR, Ramoa2025npjQI}, and quantum imaginary time evolution \cite{McArdle2019npj, Motta2020Nat, Nishi2021npj}. Each of these methods faces its own drawbacks, including classical optimization over difficult landscapes, resource-intensive tomographic techniques, or post-selection. A new class of promising optimization-free methods for preparing ground and thermal states are dissipative algorithms~\cite{Lin2024PRR, Lin2025APL, Ding2026Nat, CubittarXiv2023, HagenarXiv2025}. Dissipative algorithms simulate the physical cooling process of an open quantum system coupled to a low-temperature thermal environment. 

In classical statistical mechanics, Markov chain Monte Carlo techniques such as the Metropolis-Hastings algorithm are some of the most prominent methods for simulating thermodynamic behavior. Dissipative algorithms can be seen as an extension of these techniques to quantum systems~\cite{ChenarXiv2023, GilyenarXiv2024, Ding2025CMP, Chen2025Nat}. They have been proposed as a versatile platform for preparing many-body states on quantum computers, including fractional quantum hall states~\cite{LiuarXiv2026} and ground and excited states of chemical systems~\cite{WattsarXiv2026, LiarXiv2025, Li2025npj}. Other work has studied the role of modulated system-bath couplings~\cite{Lloyd2025PRXQ, Lloyd2026PRX}, algorithm noise tolerance~\cite{PurcellarXiv2025}, randomized measurements~\cite{LangbehnarXiv2025}, optimality and performance bounds~\cite{Rouze2026PRL, Zhan2026PRX}, and incorporating dissipative components into VQE frameworks~\cite{Ilin2025IEEE}. Recently, experimental implementations of dissipative algorithms on both superconducting~\cite{FarrellarXiv2026, GetelinaarXiv2026} and trapped-ion~\cite{Seki2026PRApp} quantum devices have emerged.         

Two key conditions that must be met for implementing ground state preparation using dissipative dynamics are selecting the system-bath interaction operators to ensure that a path exists connecting any energy eigenstates on which the initial state has nontrivial support to the target ground state, and tuning the discrete bath spectrum so that it is resonant with the corresponding energy-decreasing transitions in the system. Satisfying these conditions without relying on some a priori knowledge of the system energy spectrum while maintaining efficient algorithm convergence times is a significant challenge. In this paper, we propose an adaptive implementation of dissipative state preparation that mitigates this challenge by applying a single-shot reinforcement learning (RL) scheme that optimizes the distribution of bath frequencies and system-bath interaction operators to substantially improve convergence times. The use of adaptive features is well established for improving the resource costs of quantum simulation algorithms, yet this approach is essentially unexplored for dissipative state preparation, outside of adaptive circuit structures for entangled state preparation~\cite{Pocklington2025PRL}.  

The remainder of this paper is organized as follows. In Sec.~\ref{sec:2} we review the core principles of dissipative state preparation, as outlined in Ref.~\cite{Ding2026Nat}. In Sec.~\ref{sec:3} we detail our adaptive reinforcement learning implementation, RL-Adapt, before concluding in Sec.~\ref{sec:4}. More details about the algorithm design and implementation can be found in the Appendix.

\begin{figure*}[]
\centerline{\includegraphics[width=1.0 \linewidth]{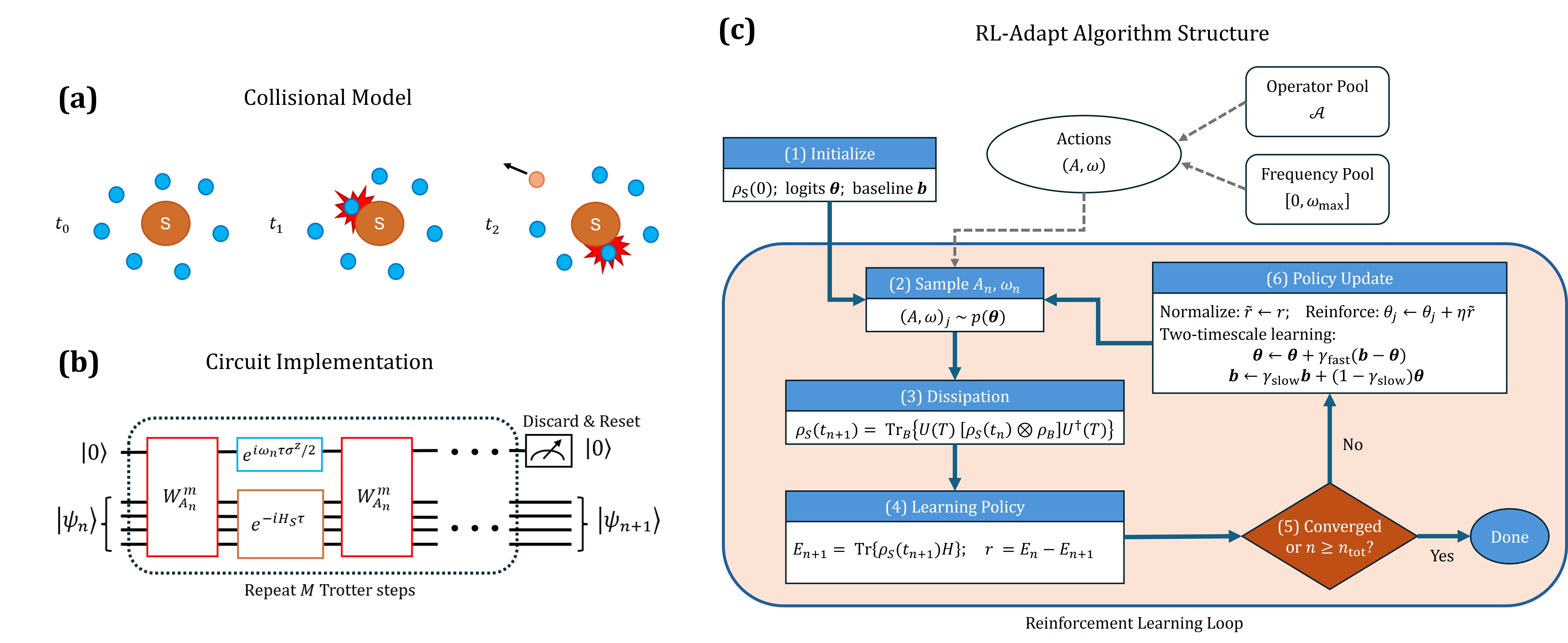}}
\caption{(a) Illustration of the collisional model of thermalization. (b) Quantum circuit for one collisional model time step of the dissipative algorithm, corresponding to step (3) of the schematic flowchart. (c) Schematic illustration of the RL-based, adaptive dissipative algorithm.}
\label{fig:Main}
\end{figure*}

\section{Dissipative State Preparation}
\label{sec:2}

Dissipative algorithms are commonly constructed as circuit-based implementations of collisional models, a microscopic open quantum systems model for thermalization widely used in quantum thermodynamics \cite{Pocrnic2025JPA, Ziman2005OSID, Strasberg2017PRX, Rodrigues2019PRL, Cattaneo2021PRL, Ciccarello2022PR, Cusumano2022Entropy}. In the collisional model, illustrated in Fig.~\ref{fig:Main}(a), a system of interest interacts with an environment composed of a large number of identical bath particles through a series of discrete interactions (``collisions"), where each interaction occurs with a fresh bath particle. One time step of the collisional model evolution is given by,
\begin{equation}
\rho_S\!\left(t_{n+1}\right)
= \operatorname{Tr}_B\!\left[
U(T)\,\bigl(\rho_S(t_n)\otimes \rho_B\bigr)\,U^\dagger(T)
\right]
\end{equation}
where 
\begin{equation}
    U(T) = e^{-\,i\bigl(H_S + H_B + H_I\bigr)\,T}
    \label{eq:FullUnitary}
\end{equation}
with $H_S$, $H_B$, and $H_I$ the Hamiltonians of the system, bath, and interaction, respectively, and $T$ is the interaction duration. We set $\hbar = 1$ throughout the remainder of the manuscript.


The dissipative algorithm simulates collisional model dynamics through a Trotterized implementation of Eq.~\eqref{eq:FullUnitary}. Each collisional model time step, $n$, consists of $M = \lceil 2 T/\tau \rceil$ Trotter steps, where $\tau$ is the interaction timescale. Each Trotter step alternates between gates that generate the system-environment interaction by coupling the system qubits to an ancilla qubit representing the bath particle,
\begin{equation}
    W^{m}_{A_n}
= \exp\!\left\{
-\,i\,\alpha\,f_m\,
\left(A_n\!\otimes\!\sigma^- + A_n^\dagger\!\otimes\!\sigma^+\right)\,
\frac{\tau}{2}
\right\}
\end{equation}
and gates implementing the system and bath free evolution,
\begin{equation}
    U_S = e^{-i H_S \tau} \quad \mathrm{and} \quad U_B = e^{i \omega_n \tau \sigma^z/2}.
\end{equation}
Here $\alpha$ is the coupling strength and $f(t) =(2\pi)^{-1/4} \sigma^{-1/2} \exp(-t^2/4 \sigma^2)$ is a time-domain Gaussian interaction envelope. For an individual Trotter time step, $m$, the interaction envelope takes the value $f_m = f\big((m + \tfrac{1}{2})\tau - T\big)$. 

At each collisional model timestep $n$, a system interaction operator $A_n$ is randomly drawn from an operator pool $\mathcal{A}$. Similarly, at each time step, the ancilla qubit energy gap $\omega_n$ is drawn from an interval $[0, \omega_{\mathrm{max}}]$ that spans the system energies, permitting an energy transition in the system of magnitude $\omega_n$. After each collisional model time step, the ancilla qubit is reset, simulating the arrival of a fresh bath particle. A circuit diagram of one time step of the dissipative model is provided in Fig.~\ref{fig:Main}(b).

Note that there are two equivalent interpretations of the role of $f(t)$. In some dissipative algorithm constructions (cf. Ref.~\cite{Lin2024PRR}), $f(t)$ is non-Gaussian and is carefully constructed such that its Fourier transform, $\hat{f}(\omega)$ plays the role of a frequency filter that suppresses undesired transitions that would increase the energy of the system. This approach lacks the independent free evolution of the bath, instead relying on a Trotterized implementation of the Lindblad jump operators. In follow-up work (cf. Ref.~\cite{Ding2026Nat}), the incorporation of separate bath evolution serves to target $f$ on the currently selected bath frequency, $\omega$. In this case, $f(t)$ instead serves as an interaction envelope whose width sets the energy window for inducing energy transitions in the system.

\section{Adaptive Operator and Bath Hamiltonian Selection}
\label{sec:3}

\subsection{Selection Criteria Trade-off}

One of the core components of the dissipative algorithm is the selection of the system-bath interaction operators and the range of ancilla energy spacing. For a successful implementation, two conditions must be fulfilled:
\begin{enumerate}[label=(\alph*)]
    \item The operator pool $\mathcal{A}$ must contain sufficient operators that a path connecting all occupied energy eigenstates in the initial state $\rho_S(t_0)$ to the ground state exists.
    \item The range of the possible ancilla energy spacings $\omega_{\mathrm{max}}$ must span the energy gaps of all the eigenstate transitions required for (a).  
\end{enumerate}
We refer to these conditions as the \textit{success conditions}, since without meeting these conditions the algorithm will converge to a steady state with energy above the ground state. There are an additional two conditions that must be fulfilled for any individual time step of the algorithm to bring the current state $\rho_S(t_n)$ closer to the ground state:
\begin{enumerate}[label=(\alph*), resume]
    \item A transition matrix element for the selected operator, $\langle \psi_u \vert A_n \vert \psi_v \rangle$, where $E_u < E_v$, must be non-zero for an energy eigenstate $\vert \psi_v \rangle$ that contributes to $\rho_S(t_n)$.    
    \item The selected ancilla energy spacing $\omega_n$ must be resonant with the corresponding transition energy, $\Delta_{vu} = E_v - E_u$ within the tolerance of the interaction envelope.  
\end{enumerate}
We refer to (c) and (d) as the \textit{efficiency conditions}, since if they are fulfilled for each time step there will be few ``failed" interactions that do not decrease the system energy, improving the algorithm convergence time and reducing the necessary circuit depth.    

There is an inherent tension between optimizing $\mathcal{A}$ and $\omega_{\mathrm{max}}$ for both the success and efficiency conditions simultaneously. Let us first consider the choice of operator pool $\mathcal{A}$. In order to give the best chance of fulfilling condition (a), we want a large operator pool. In the most extreme case, choosing an operator pool consisting of all possible Pauli strings of length $d_S$, where $2^{d_S}$ is the dimension of the Hilbert space of the system Hamiltonian, would ensure that the full Hilbert space is accessible from an arbitrary initial state, guaranteeing a path to the ground state. However, whether the interaction operators are applied randomly or sequentially from the pool, such a large operator pool decreases the likelihood that condition (c) is fulfilled for a particular time step, especially during the later stages of the algorithm, when much of the occupation has already concentrated in low-energy eigenstates.

An analogous argument can be made for a tension between conditions (b) and (d). Since we must assume little or no a priori knowledge of the spectrum of $H_S$, our best chance of fulfilling condition (b) is to choose a large value of $\omega_{\mathrm{max}}$ to ensure that the range of ancilla energy spacings covers all possible transition energies. However, exactly the same issue arises, a large value of $\omega_{\mathrm{max}}$ decreases the likelihood that condition (d) is fulfilled for a particular time step as the algorithm progresses and fewer transitions are resonant for the current density matrix. To resolve this tension, we implement an adaptive approach to the dissipative algorithm that adjusts the selection probabilities for $A_n$ and $\omega_n$ to favor operators that provide large energy decreases using single-shot reinforcement learning.

Before we describe the implementation, let us first consider the key features that we want the RL algorithm to learn. First, we want it to learn to preference values of $\omega$ resonant with the system energy eigenstate transitions. In other words, the initially flat $\omega$ distribution should adjust to encode the information about the spectrum of $H_S$. Second, we want it to learn to favor pairing each $A$ operator with the energy eigenstate transitions induced by that operator. For example, let us imagine a particular operator $A_i$ has non-zero transition matrix elements at $\langle \psi_1 \vert A_i \vert \psi_3 \rangle$ and $\langle \psi_2 \vert A_i \vert \psi_5 \rangle$. We want the algorithm to learn to preference $\omega \approx \Delta_{3,1}$ and $\omega \approx \Delta_{5,2}$ on steps when $A_i$ is selected from the pool.

Both of these features are independent of the system state, $\rho_S(t_n)$, and are relevant at any time step in the algorithm. However, an algorithm that bases the probability of an $(A,\omega)$ action selection only on these two features faces a potential flaw. Consider the case in which an early choice of an $(A,\omega)$ pairing shifts a significant population from a highly excited state to a low-energy state, producing a large decrease in the energy. Consequently, the RL will favor this action going forward, even after the point that the population in the excited state has been exhausted and this particular pair is no longer useful. Thus, the third feature that the algorithm should learn is, among the favored $(A,\omega)$ actions where the $\omega$ value matches a transition generated by $A$, to further favor actions corresponding to ``live" transitions for the current state $\rho_S(t_n)$. For a transition $\ket{\psi_v} \rightarrow \ket{\psi_u}$ to be ``live" we mean that $\rho_S(t_n)$ has nonzero population in $\ket{\psi_v}$.

\subsection{Two-timescale RL Learning for Operator and Frequency Selection}

To accomplish these goals, we implement a two-timescale learning rule. A slow, long-memory ``baseline" residual gradually learns the spectral data over the course of the full algorithm. A fast ``active" residual learns the live transitions relevant for the current system state, but has a short memory, quickly reverting back to the baseline established by the slow residual.      

In the rest of this section, we describe the key aspects of our implementation of the dissipative algorithm with RL, with a detailed breakdown of the full algorithm and additional details provided in Appendix~\ref{sec:AppendixA}. A schematic illustration of the main algorithm logic is provided in Fig.~\ref{fig:Main}(c).  

We begin by discretizing the range of possible ancilla energies into $\omega_{\mathrm{max}}/N_{\omega}$ equally sized bins. We then initialize a zero array, $\bm{\theta}$, of $N_{\mathcal{A}} \times N_{\omega}$ logits, each corresponding to a possible $(A,\omega)$ pair. We similarly initialize an equally sized baseline array, $\bm{b}$, that will serve to track the contribution of the slow-timescale learning to the logits. The basis of our reward function is the decrease in energy expectation value at each step $r_n=\mathrm{Tr}(H\rho_n)-\mathrm{Tr}(H\rho_{n+1})$. Since the magnitude of each energy drop will naturally decrease as the system converges towards the ground state, to avoid biasing the learning towards large early drops we calculate moving estimates of the reward mean and variance with a decay factor $\delta$,   
\begin{align}
    \text{Mean:} \  \bar r &\leftarrow \delta \bar r+(1-\delta) r_n, \\
    \text{Variance:} \ v & \leftarrow \delta v+(1-\delta)(r_n-\bar r)^2
\end{align}
and use the standardized reward as the actual learning signal,
\begin{equation}
    \tilde r_n=\frac{r_n-\bar r}{\sqrt{v}}.
\end{equation}
The learning policy is a softmax over the logit $\theta_{j}$ with a learning temperature $\Theta_l$,
\begin{equation}
P(A,\omega)=\frac{\exp(\theta_j/\Theta_l)}{\sum_{j}\exp(\theta_j/\Theta_l)}
\end{equation}
After each time step, the chosen logit is updated based on the standardized reward with a learning rate of $\eta$,
\begin{equation}
    \theta_j \leftarrow \theta_j+\eta\,\tilde r_n.
\end{equation}
The complete logit array is then updated with two additional coupled operations, implementing the two-timescale learning rule. The fast residual is updated via a baseline reversion with a rate $\gamma_{\mathrm{fast}}$,  
\begin{equation}
    \bm{\theta} \leftarrow \bm{\theta}+\gamma_{\mathrm{fast}}(\bm{b}-\bm{\theta}),
\end{equation}
where $\bm{b}$ is the baseline at the current step.
The baseline is then updated via an exponential moving average with a rate $\gamma_{\mathrm{slow}}$, 
\begin{equation}
    \bm{b} \leftarrow \gamma_{\mathrm{slow}}\bm{b}+(1-\gamma_{\mathrm{slow}}) \bm{\theta}.
    \label{eq:baseline}
\end{equation}

It is important to highlight that this is a \textit{single-shot} RL scheme, and does not rely on any information learned in previous implementations. This is by necessity, as the goal of the RL is to speed up the convergence to the target ground state. If the ground state was previously known from an earlier implementation of the algorithm, however inefficient, it would not be necessary to run the algorithm in the first place. In addition, the two-timescale learning scheme we adopted here does not increase the algorithm's quantum resource cost compared to the single-shot single-timescale RL schemes. Instead, the computational complexity is completely classical. 

\section{Implementations}

To verify the capabilities of the RL-Adapt dissipative algorithm, we implement it on both spin Hamiltonians, in the form of the one-dimensional XY-model, and a range of electronic structure Hamiltonians. For each of our benchmark tests, we choose $\omega_{\mathrm{max}}$ to be equal to the magnitude of the largest system energy gap.     

\subsection{XY-Model}

\begin{figure}
	\subfigure[$N = 4$]{
            \includegraphics[width=.22\textwidth]{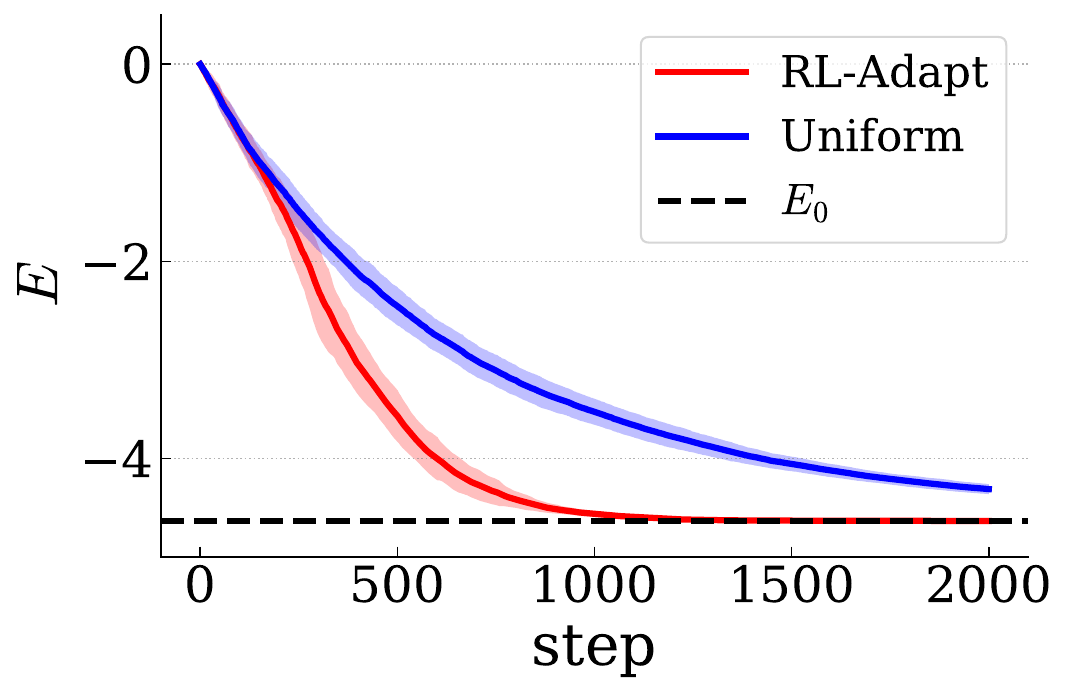}
	}
	\subfigure[$N = 6$]{
		\includegraphics[width=.22\textwidth]{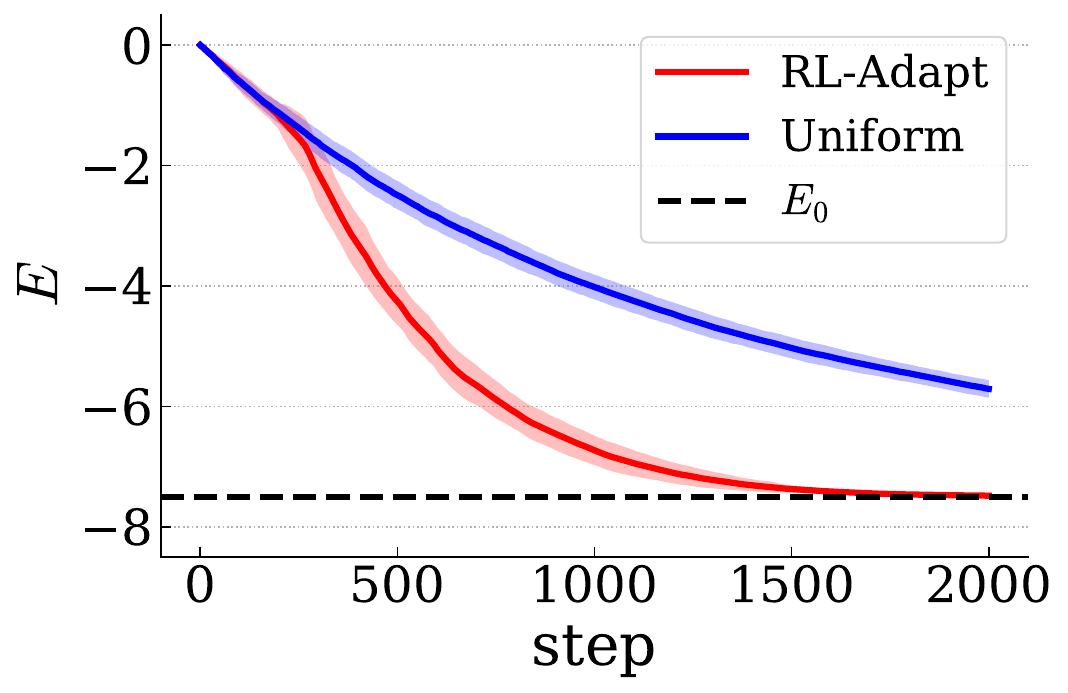}
	}
	\caption{\label{fig:XYModel}Energy as a function of time for the two-timescale reinforcement learning dissipative algorithm applied to the 1D XY model with (a) 4 sites and (b) 6 sites. Plotted curves are averages over 50 realizations of the algorithm, with the shaded regions corresponding to one standard deviation. Hamiltonian parameters are $J = 1$, $h = -0.7$. RL parameters are $n_{\omega} = 40$, $\eta = 2.0$, $\gamma_{\mathrm{fast}} = 40/n$ and $\gamma_{\mathrm{slow}} = 2/n$.}
\end{figure}

In one dimension, the Hamiltonian for the XY-model in an external field is,
\begin{equation}
    H = -J \sum_{i=1}^{N-1} \left(\sigma_i^x \sigma_{i+1}^x + \sigma_i^y \sigma_{i+1}^y \right) - h \sum_{i=1}^N \sigma_i^z
\end{equation}
In Fig.~\ref{fig:XYModel} we plot the convergence of the energy to the ground state of the 1D XY-model for system sizes of $N=4$ and $N=6$ with open boundary conditions. We choose the maximally mixed state, corresponding to the system at infinite temperature, as our initial state. This choice serves as a strong test of the algorithm's ability to explore the system Hilbert space, since it requires that $\mathcal{A}$ can generate a viable path connecting every energy eigenstate to the ground state. For our operator pool, we choose $\mathcal{A}$ to contain all single-site Pauli operators, $A = \sigma^{\alpha}_{j}$, $\alpha \in \{x,y,z\}$, $j \in \{1,\, ...\, , N\}$. This operator pool provides a balance between ergodicity and size, scaling linearly with the system dimension.  

As a benchmark comparison, we compare the performance of the RL-Adapt dissipative algorithm (red lines in Fig.~\ref{fig:XYModel}) with a non-adaptive implementation where $(A,\omega)$ are chosen randomly from a uniform distribution at each time step (blue lines in Fig.~\ref{fig:XYModel}). Since the algorithm is inherently stochastic, we average the energy convergence over 50 independent shots of both the RL-Adapt and non-adaptive implementations to ensure that any advantage in convergence speed is not just a result of a particularly successful individual run. We see that for both system sizes, the RL-Adapt approach shows significantly faster convergence to the ground state of the model (black dashed line in Fig.~\ref{fig:XYModel}). 

\begin{figure}[]
\centerline{\includegraphics[width=0.9 \linewidth]{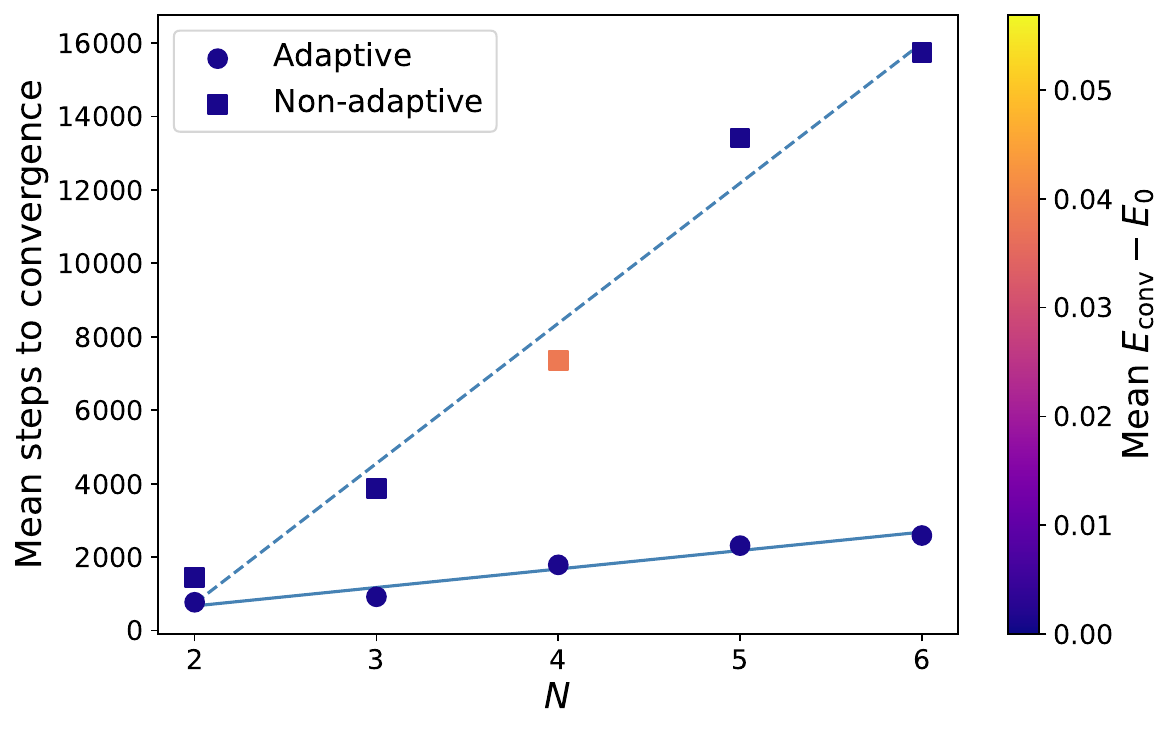}}
\caption{Average steps to ground state convergence as a function of system size for the 1D XY-model. Solid and dashed lines are linear best fits for the RL-Adapt and non-adaptive implementations respectively. The color scale gives the mean error in the converged ground state energy. Hamiltonian parameters are $J = 1$, $h = -0.7$. RL parameters are $n_{\omega} = 40$, $\eta = 2.0$, $\gamma_{\mathrm{fast}} = 40/n_{\mathrm{tot}}$ and $\gamma_{\mathrm{slow}} = 2/n_{\mathrm{tot}}$.}
\label{fig:XYConv}
\end{figure}

To quantify the enhancement in convergence time, in Fig.~\ref{fig:XYConv} we plot the convergence time as a function of system size, $N$. For these benchmark tests, convergence is defined by reaching an error threshold of $10^{-3}$ from the true ground state, or meeting a stall criterion of failing to produce an energy decrease greater than $10^{-4}$ over 1000 consecutive time steps. Using a linear fit, we find an approximate relationship of $\sim 3800/N$ steps to convergence for the non-adaptive implementation and $\sim 500/N$ steps to convergence for the RL-Adapt implementation, representing a $7.6$-fold enhancement in convergence time. The $N=4$ non-adaptive data point merits further discussion due to its failure to converge to the true ground state. Instead, the fixed point of the channel generated by $U(T)$ with a flat distribution for $P(A,\omega)$ is a steady state consisting of approximately 80\% of the population in the ground state and 20\% of the population in the first excited state\footnote{In the single ancilla dissipative ground state algorithm formulation in Ref.~\cite{Lin2024PRR} $f(t)$ serves as a filter function to suppress energy excitations in the system (though excitations can still occur, if the tails of the filter function in frequency space extend into the positive domain). Here, $f(t)$ instead defines a frequency interaction window. Due to the width of the Gaussian and the choice of frequency spacings, the dissipator can produce excitations in the system.}. Ultimately, this is a consequence of the fact that the choice of $N_{\omega} = 40$ equally spaced $\omega$ values happens to result in no value of $\omega$ that is close to the ground state energy gap for $N=4$. Thus, the non-adaptive channel is only weakly resonant with this transition and never fully converges to the ground state. Despite the same $\omega$ discretization in the adaptive case, the biased distribution is able to sufficiently prioritize the $(A,\omega)$ actions that are weakly resonant with this transition while disfavoring other actions that generate fluctuations of population back out of the ground state, resulting in ground state convergence. This highlights another advantage of the RL-Adapt implementation, namely that it can produce convergence, even for coarse $\omega$ discretizations that fail to converge in a non-adaptive implementation.     

\begin{figure}
	\subfigure[]{
            \includegraphics[width=.38\textwidth]{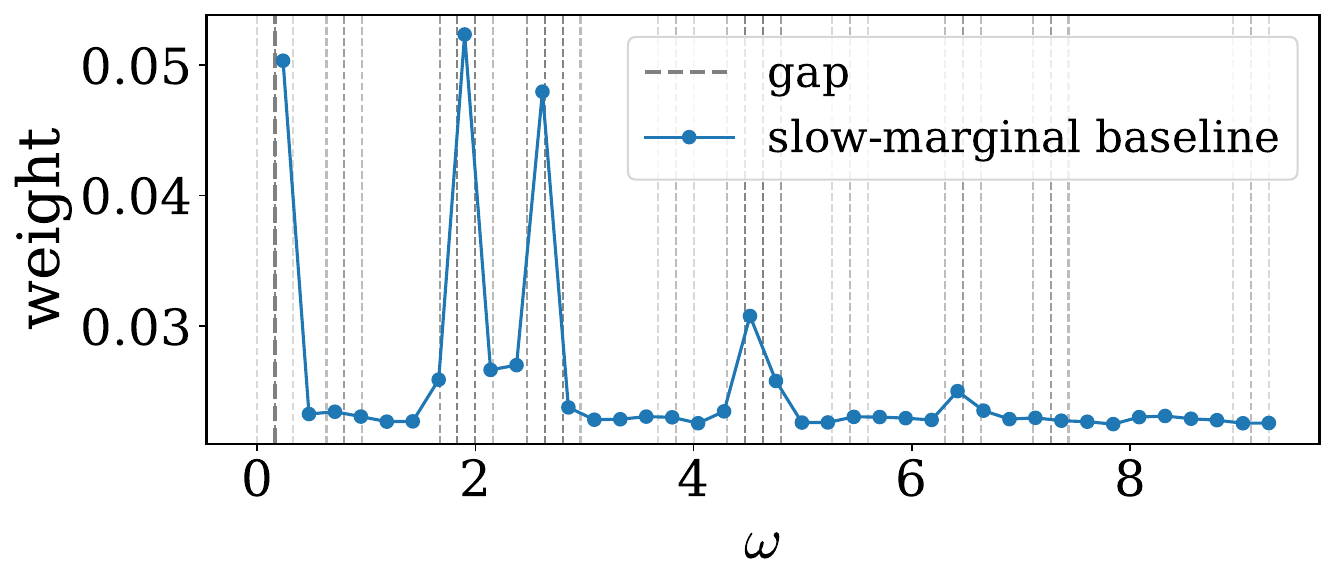}
	}
	\subfigure[]{
		\includegraphics[width=.38\textwidth]{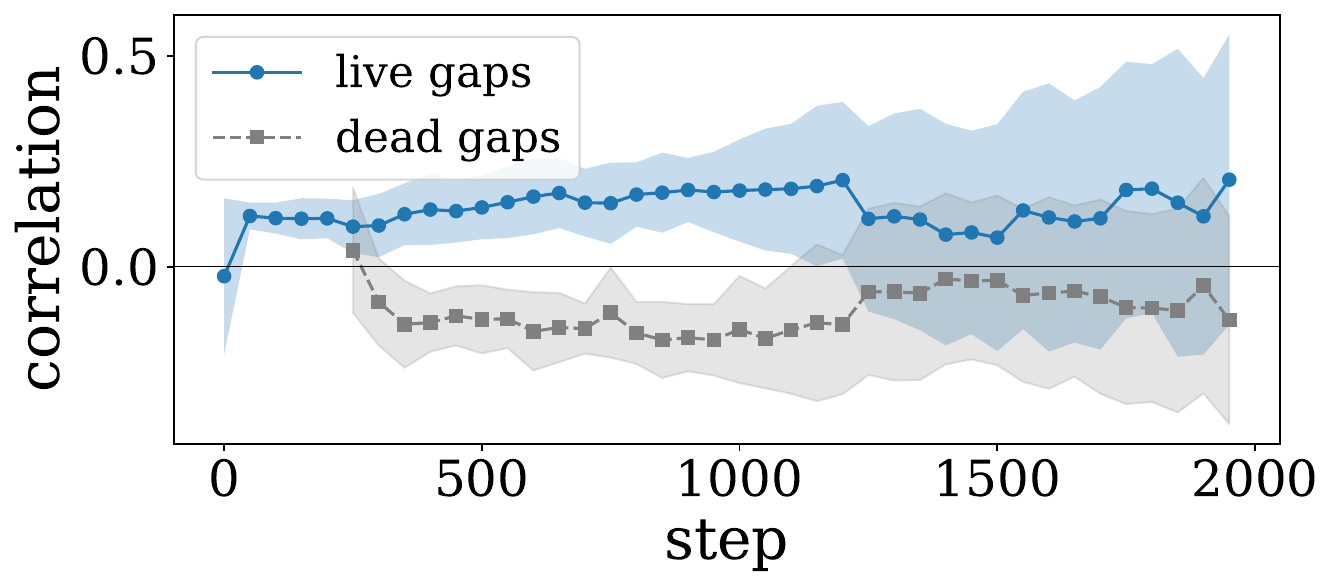}
	}
	\caption{\label{fig:XYDiag} (a) Average final $\omega$ distribution weights for the slow baseline marginal overlaid with the spectral gaps of $H_S$ for the XY-model with $N=4$. The blue dots show the weights from the learned, slow-timescale RL baseline ($\bm{b}$ in Eq.~\eqref{eq:baseline}) for choosing the corresponding $\omega$, while the dashed lines show energy gaps of the problem Hamiltonian, $\Delta_{u, v}$ for all energy eigenstates $\ket{\psi_u}$ and $\ket{\psi_v}$. (b) Average correlation between $\omega$ values with above-baseline weight and gaps corresponding to feasible ``live" (blue) or infeasible ``dead" (gray) transitions in $\rho_S$ as a function of time step. The shaded region corresponds to one standard deviation. In both panels, averages are taken over 50 shots of the algorithm. Hamiltonian parameters are $J = 1$, $h = -0.7$. RL parameters are $n_{\omega} = 40$, $\eta = 2.0$, $\gamma_{\mathrm{fast}} = 40/n_{\mathrm{tot}}$ and $\gamma_{\mathrm{slow}} = 2/n_{\mathrm{tot}}$.}
\end{figure}

In order to verify that the two-timescale learning rule is functioning as expected, in Fig.~\ref{fig:XYDiag}(a), we compare the average final $\omega$ distribution weights for the slow baseline marginal to the real spectral gaps for the four-site XY-model. We see that the $\omega$ values with higher weight align directly with intervals containing a high density of spectral gaps, indicating that the baseline is in fact learning the spectrum of $H_S$. In Fig.~\ref{fig:XYDiag}(b), we plot the average Pearson correlation coefficient between the favored (above baseline weight) $\omega$ values at step $i$ and the ``live" gaps, corresponding to all transitions from states with non-zero weight in $\rho_S(t_n)$. Details of how the correlation is calculated are provided in Appendix~\ref{sec:AppendixC}. We see that there is a consistent positive correlation, indicating that the fast residual is correctly learning to favor live gaps. To ensure that it is favoring \textit{only} the live gaps, and not just favoring all real gaps, we also plot the correlation with the dead gaps, corresponding to transitions from states with no population in $\rho_S(t_n)$. We see that the fast residual is consistently negatively correlated with these gaps, demonstrating that the fast residual really is singling out only the live transitions. The shaded regions around each correlation line show one standard deviation. We observe that at later times the magnitude of both the positive and negative correlations drop, and the shot-by-shot fluctuations increase. This is a consequence of the fact that as the state converges to the ground state, fewer and fewer live gaps exist. Consequently, the exploration of the possible transition energies induced by the short memory of the fast residual is less and less likely to find new live gaps, degrading the correlation. We note that since the $\rho_S(t_0)$ is taken to be the infinite temperature state with equal population in each eigenstate, initially there are no dead gaps. This accounts for the delay in the onset of the dead gap correlation in Fig.~\ref{fig:XYDiag}(b).  

These metrics have shown that the RL scheme is successfully learning to prioritize $\omega$ values corresponding to available eigenstate transitions in $\rho_S$. In the following, we are going to verify that the RL scheme is also learning to pair these $\omega$ values with the system interaction operators from $\mathcal{A}$ that generate those transitions. We will refer to a logit $(A,\omega)$ pair as ``aligned" if the $\omega$ bin corresponds to the energy of a transition with a non-zero element in the transition matrix generated by $A$. For each operator in $\mathcal{A}$ we examine what fraction of the total probability mass for that operator is assigned to aligned pairs. We define the enrichment of that operator as the factor by which the probability mass assigned to aligned pairs exceeds the uniform distribution expected from totally random probability mass assignment. High enrichment indicates that the learned slow baseline preferentially concentrates weight on frequencies corresponding to transitions that can be driven by the associated interaction operator. Details of the enrichment calculation are provided in Appendix~\ref{sec:AppendixD}. In Fig.~\ref{fig:XYEnrichment} we plot the enrichment for each of the 12 single-site Pauli operators in $\mathcal{A}$ for the 4-site XY-model. We see that the average raw enrichment for each operator is greater than one, confirming that the algorithm is learning to favor aligned $(A,\omega)$ choices. The point color in Fig.~\ref{fig:XYEnrichment} indicates the number of eigenstate transitions induced by that operator. We note that the raw enrichment is the largest for operators with an intermediate number of induced transitions. Naively, we would expect operators that induce many transitions to have the greatest enrichment since they should have more aligned pairs and should receive the most positive reinforcement. However, this behavior is actually a consequence of the fact that the enrichment has a ceiling given by $N_{\omega}/N_{\mathrm{target}}$, where $N_{\mathrm{target}}$ is the number of target $\omega$ bins corresponding to the energy of an aligned transition. Operators that induce a large number of transitions are likely to have a large $N_{\mathrm{target}}$, leading to a lower enrichment ceiling. To account for this, in the inset of Fig.~\ref{fig:XYEnrichment} we plot the enrichment normalized by the ceiling to determine $f_{\mathrm{max}}$, the fraction of the maximum possible enrichment achieved by each operator. We see the expected behavior emerge under this normalized metric, with the high transition number operators achieving the largest pair alignment, at around $70\%$ of the maximum enrichment.      

\begin{figure}[]
\centerline{\includegraphics[width=0.9 \linewidth]{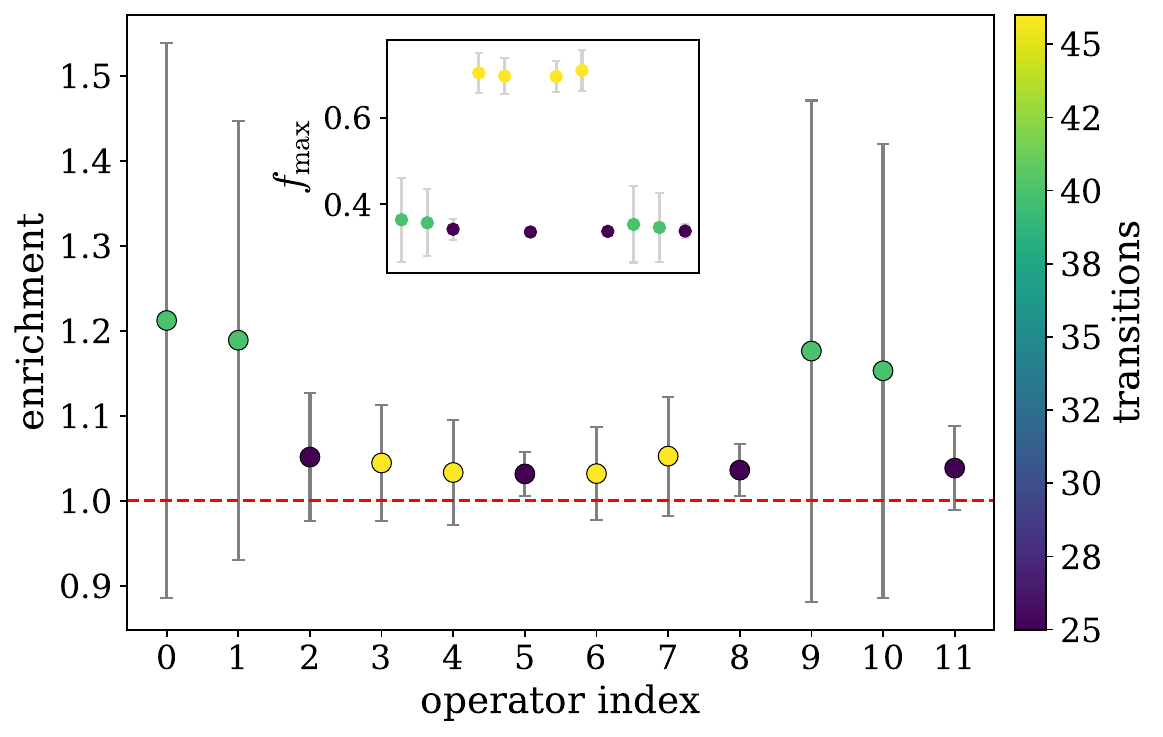}}
\caption{Average enrichment (defined as the factor by which the probability mass assigned to aligned pairs exceeds the uniform distribution) over 50 shots of the RL-Adapt dissipative algorithm for the 4-site XY model with an operator pool, $\mathcal{A}$, consisting of each single-qubit Pauli operator. The color scale indicates the number of eigenstate transitions induced by each operator. The dashed red line indicates the baseline enrichment expected from a uniform assignment of probability mass. The error bars indicate one standard deviation. The inset plot provides the enrichment normalized by the enrichment ceiling $N_{\omega}/N_{\mathrm{target}}$. Hamiltonian parameters are $J = 1$, $h = -0.7$. RL parameters are $n_{\omega} = 40$, $\eta = 2.0$, $\gamma_{\mathrm{fast}} = 40/n_{\mathrm{tot}}$ and $\gamma_{\mathrm{slow}} = 2/n_{\mathrm{tot}}$.}
\label{fig:XYEnrichment}
\end{figure}

\subsection{Electronic Structure Hamiltonians}

To demonstrate the performance of the adaptive dissipative algorithm beyond spin chain systems, we also apply it to a range of electronic structure Hamiltonians. To illustrate the ability of the RL-Adapt dissipative algorithm to capture the correlation energy, for the following examples we choose the initial system state to be the Hartree-Fock state. We build $\mathcal{A}$ from all fermionic single and double excitation operators. 

\begin{figure}
	\subfigure[$\mathrm{H}_2$]{
            \includegraphics[width=.22\textwidth]{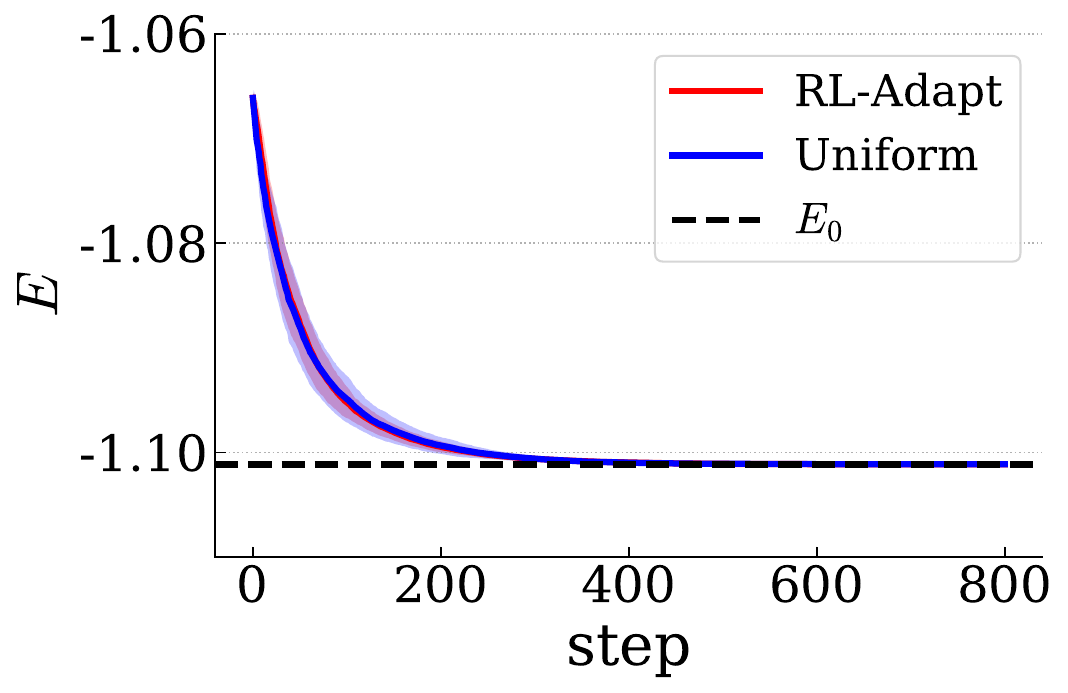}
	}
	\subfigure[$\mathrm{H}_4$]{
		\includegraphics[width=.22\textwidth]{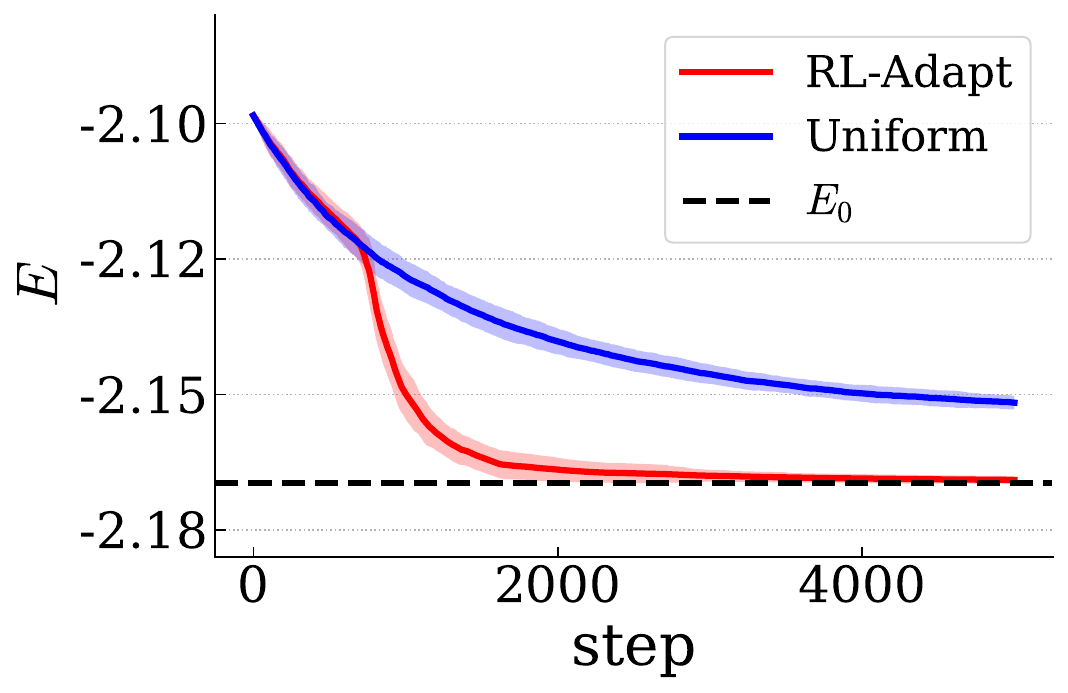}
	}
	\caption{\label{fig:HChem}Energy as a function of time for the two-timescale reinforcement learning dissipative algorithm applied to (a) $\mathrm{H}_2$ and (b) $\mathrm{H}_4$ with a $1 \AA$ separation using the STO-3G basis set. The $\mathrm{H}_4$ molecule is in a linear configuration. Plotted curves are averages over 50 realizations of the algorithm, with the shaded regions corresponding to one standard deviation. RL parameters are $n_{\omega} = 80$, $\eta = 2.0$, $\gamma_{\mathrm{fast}} = 40/n_{\mathrm{tot}}$ and $\gamma_{\mathrm{slow}} = 2/n_{\mathrm{tot}}$.}
\end{figure}

In Fig.~\ref{fig:HChem} we plot the energy convergence for $\mathrm{H}_2$ and  $\mathrm{H}_4$ Hamiltonians. We see that the performance of the adaptive and non-adaptive implementations is nearly identical for $\mathrm{H}_2$. This is a consequence of the fact that the operator pool of singles and doubles excitations for $\mathrm{H}_2$ consists of only three operators. With such a small operator pool, the RL scheme contributes very little to the operator selection. In contrast, for $\mathrm{H}_4$, where the operator pool contains 26 excitation operators, we see a significant advantage in convergence time emerge for the adaptive implementation.     

For a larger system size of 10 qubits, we test the RL-Adapt algorithm on a reduced-dimensionality effective Hamiltonian for LiH (using the cc-pVTZ basis from the original basis set exchange) generated by coupled-cluster downfolding techniques~\cite{Bauman2019JCP, Kowalski2020JCP, Bauman2022JCP, Huang2023PRXQ, Bauman2026PRR}. The effective Hamiltonian considered here is generated using the double unitary coupled cluster (DUCC) ansatz, which expresses the exact wave function as a product of two unitary exponential ans\"atze, one defined by a cluster operator carrying only spin orbital indices corresponding to a reduced-dimensionality active space, and the other carrying at least one external spin orbital index from outside the active space. Using the DUCC ansatz, an effective Hamiltonian can be constructed in the active space whose ground state energy is identical to that of the original Hamiltonian (subject to approximations in the representation of the external cluster operator).

\begin{figure}
	\subfigure[LiH (Bare)]{
            \includegraphics[width=.22\textwidth]{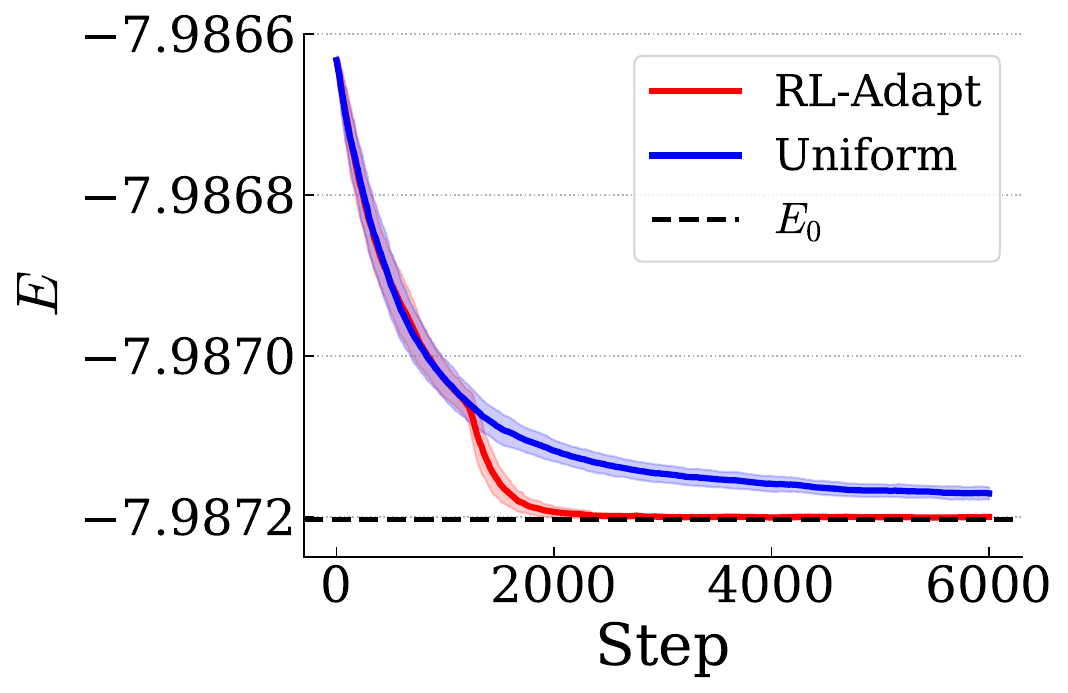}
	}
	\subfigure[LiH (DUCC)]{
		\includegraphics[width=.22\textwidth]{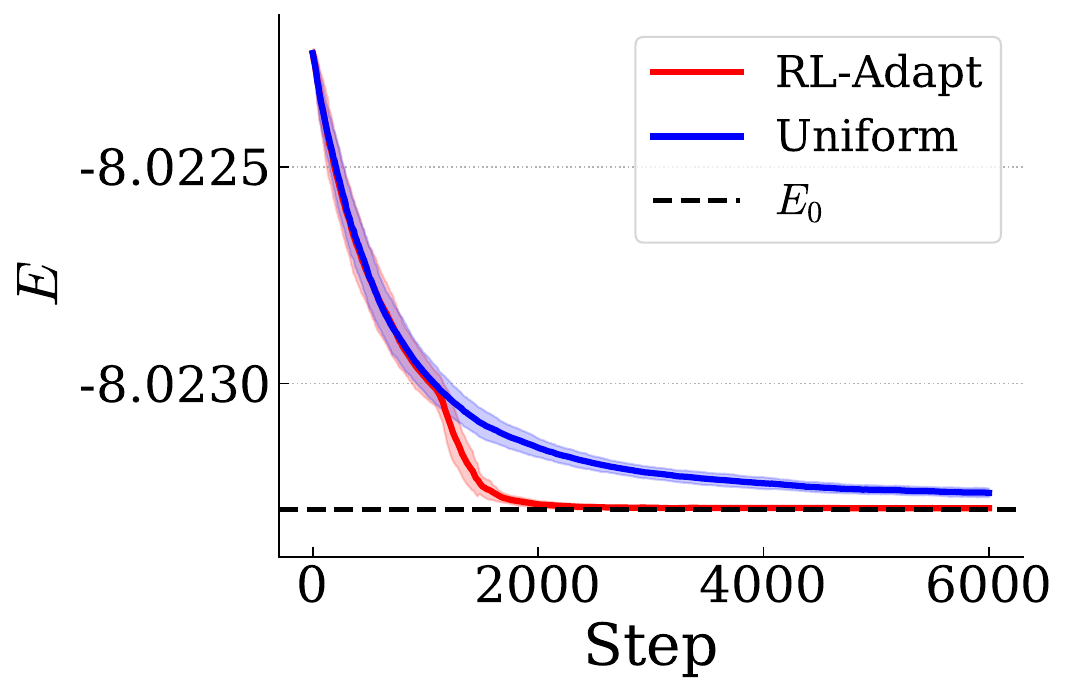}
	}
	\caption{\label{fig:LiHChem}Energy as a function of time for the RL-Adapt dissipative algorithm applied to the (a) bare and (b) downfolded DUCC Hamiltonian for LiH using the cc-pVTZ basis set in a 5-orbital (10-qubit) active space. Plotted curves are averages over 50 realizations of the algorithm, with the shaded regions corresponding to one standard deviation. RL parameters are $n_{\omega} = 80$, $\eta = 2.0$, $\gamma_{\mathrm{fast}} = 40/n_{\mathrm{tot}}$ and $\gamma_{\mathrm{slow}} = 2/n_{\mathrm{tot}}$.}
\end{figure}

In Fig.~\ref{fig:LiHChem} we plot the energy dissipation for the bare Hamiltonian (consisting of only the active space orbitals with no downfolding applied) and the DUCC downfolded Hamiltonian. We see that in both cases, the RL-Adapt algorithm provides a similar advantage in convergence time over the non-adaptive implementation. While downfolding does not change the operator structure or ground state, the general Hamiltonian spectrum is not preserved. These results indicate that the spectral changes from downfolding do not qualitatively hinder or enhance the adaptive performance for this example case. We note that, as expected, the ground state of the downfolded Hamiltonian is lower than that of the bare Hamiltonian, as a result of the additional correlation energy arising from out of active space correlations accounted for by the downfolding procedure. 

\section{Concluding Remarks} 
\label{sec:4}

There is a well-developed tradition of optimizing the resource costs of quantum simulation algorithms by introducing adaptive features. Here we extend this tradition to dissipative state preparation. With that said, there are some notable differences between the RL-Adapt dissipative algorithm detailed here and other adaptive methods. Typically, adaptive quantum simulation algorithms iteratively build an ansatz for the target state that is expanded with operators from an established pool with the goal of minimizing an objective function, such as the energy in case of ADAPT-VQE~\cite{Grimsley2019Nat}, or a distance measure to the target state in the case of adaptive variational quantum dynamics~\cite{Yao2021}. The dissipative algorithm, however, is fundamentally an open quantum system simulation and necessitates a different approach. Instead of adaptively building an ansatz state from a static operator pool, a broad pool of interaction operator-bath frequency pairings is systematically reduced to focus only on the optimal operator-frequency pairs. This approach has a distinct advantage in that it does not require the initial ansatz state to have any overlap with the target ground state.

A persistent challenge in the implementation of dissipative algorithms is that selecting the optimal pool of bath interaction operators and ancilla frequency range requires some information about the system spectrum. The dependence on a priori knowledge can be reduced by broadening the operator pool to cover as much of the system Hilbert space as possible and by expanding the frequency range, but this results in long convergence times and costly circuit depths. The RL-Adapt dissipative algorithm overcomes this issue by incorporating the spectral information encoded in each discrete energy decrease of the dissipative dynamics through reinforcement learning.

To ensure that the RL scheme learns both the time-independent spectrum of the system Hamiltonian as well as how to prioritize aligned interaction operator-bath frequency pairings that provide a strong dissipation channel for the current system state we constructed a two-timescale learning policy with a long memory for the spectral information and a short memory for the active transitions. We benchmarked the algorithm performance for the one-dimensional XY-model spin chain at a range of system sizes, showing that the adaptive model yields between a 7- and 8-fold improvement in convergence time over a non-adaptive implementation with a uniform $(A,\omega)$ distribution. We also found that the RL-Adapt implementation successfully produced convergence for coarse $\omega$ discretizations that fail to converge for the non-adaptive implementation. We further applied the RL-Adapt dissipative algorithm to a range of electronic structure Hamiltonians, demonstrating that the convergence rate enhancement also holds for chemical systems.

The RL-Adapt dissipative algorithm offers a foundation for a wide range of future research directions including further optimizations to the RL framework, applications to thermal state preparation, and alternative reward functions that prioritize other dynamical behaviors, such as minimizing mixing time.    

\begin{acknowledgments}

The authors gratefully acknowledge fruitful discussions with Ayush Asthana, Linghua Zhu, Yanzhu Chen, and Nishchay Suri in the early stages of this work. NMM and KK acknowledge support from the ``Embedding QC into Many-body Frameworks for Strongly Correlated Molecular and Materials Systems''  project, which is funded by the U.S. Department of Energy, Office of Science, Office of Basic Energy Sciences, the Division of Chemical Sciences, Geosciences, and Biosciences (under FWP 72689). 
CL, NPB, and KK acknowledge support from Pacific Northwest National Laboratory’s Quantum Algorithms and Architecture for Domain Science (QuAADS) Laboratory Directed Research and Development (LDRD) Initiative. 
This research used resources of the National Energy Research Scientific Computing Center (NERSC), a DOE Office of Science User Facility supported by the Office of Science of the U.S. Department of Energy under Contract No. DE-AC02-05CH11231 using NERSC award  NERSC DDR-ERCAP0038957.
The Pacific Northwest National Laboratory is operated by Battelle for the U.S. Department of Energy under Contract DE-AC05-76RL01830.

\end{acknowledgments}
\hfill \break
\appendix

\onecolumngrid

\section{Algorithm Structure and Methodology}
\label{sec:AppendixA}

In this appendix we provide detailed pseudocode for our implementation of the RL-Adapt dissipative algorithm in Algorithm~\ref{LearningAlgo}. We also provide some comments on our methodology.

\textit{Precomputation of Unitaries --} If the total number of $(A,\omega)$ pairs, $N_A \times N_{\omega}$, is comparable to the total number of time steps, $n_{\mathrm{tot}}$, it can be more efficient to precompute a database of all possible system-bath evolution unitaries, $U(T)$. The dynamical portion of the algorithm then only needs to apply the precomputed unitary matrix. If $N_A \times N_{\omega} \gg n_{\mathrm{tot}}$ then it is likely more efficient to compute $U(T)$ individually at each time step for the chosen $(A,\omega)$ pair, as it is likely that many potential pairings will never be selected by the algorithm, making the effort to compute the corresponding unitaries wasted. For further optimization, a hybrid approach can be taken, where the first time a particular $(A, \omega)$ pair is selected, the algorithm calculates the corresponding $U(T)$ and stores it for potential future re-use. 

\textit{Learning temperature and Annealing --} Reinforcement learning schemes typically incorporate a learning temperature, $T_l$, that is gradually reduced as the algorithm progresses. This simulated annealing promotes early exploration of the action space at high temperatures, with the cooling promoting further concentration on the optimal actions as the algorithm progresses. In our RL scheme we instead choose a fixed temperature. Since the optimal actions are not static, but depend on the current system state, annealing leads to focusing on actions that are useful for early system states but useless later on when the population in the corresponding energy eigenstates is exhausted. A fixed temperature promotes some degree of exploration at each time step so that the algorithm continuously seeks the evolving optimal actions. We note that the two-timescale learning scheme captures similar benefits to simulated annealing in RL with a static distributions by distinguishing between the static (Hamiltonian spectral gaps) and dynamic (current ``live" gaps) contributions to the optimal action distribution.

\begin{algorithm}[]
\caption{RL-Adapt Dissipative Ground State Preparation}
\SetKwInput{KwInput}{Input}
\KwInput{Initial system state: $\rho_0$, system Hamiltonian: $H_S$, maximum ancilla energy gap: $\omega_{\mathrm{max}}$, number of $\omega$ bins: $N_{\omega}$, system interaction operator pool: $\mathcal{A}$ (of size $N_A$), interaction envelope function: $f(t)$, Trotter time step: $\tau$, interaction duration: $T$, coupling strength: $\alpha$, number of interactions: $n_{\mathrm{tot}}$.}
\KwInput{Learning rate: $\eta$, reward decay: $\delta$, RL temperature: $\Theta_l$, fast reversion rate: $\gamma_{\mathrm{fast}}$, slow baseline rate: $\gamma_{\mathrm{slow}}$.}

\SetKwFunction{FMain}{Dissipation}
\SetKwProg{Fn}{Function}{:}{\KwRet}

\SetKwFunction{FAncilla}{PrecomputeAncillaUnitaries}
\Fn{\FAncilla{$\omega_{\mathrm{max}}$, $N_{\omega}$, $\tau$}}{
    Initialize \textit{UBArray}\;
    \For{$j = 1$ \KwTo $N_{\omega}$} {
        $\omega_j = j\,(\omega_{\mathrm{max}}/N_{\omega})$\;
        $U_B^{(j)} = \cos(\omega_j \tau/2)\,I + i \sin(\omega_j \tau/2)\,\sigma^Z$\;
        Append $U_B^{(j)}$ to \textit{UBArray}\;
    }
    \KwRet \textit{UBArray}\;
}

\SetKwFunction{FTot}{PrecomputeTotalUnitaries}
\Fn{\FTot{\textit{UBArray}, $U_S$, $\mathcal{A}$, $N_{\omega}$, $f(t)$, $\tau$, $T$, $M$, $\alpha$}}{
    Initialize \textit{UTotArray}\;
    \For{$j = 1$ \KwTo $N_{\omega}$} {
        $U_B^{(j)} = \textit{UBArray}[j]$\;
        \For{$k = 1$ \KwTo $N_A$}{
            $A_k = \mathcal{A}[k]$\;
            $U_{\mathrm{tot}}^{(j,k)} \leftarrow I$\;
            \For{$l = 1$ \KwTo $M$} {
                $f_l = f\big((l + \tfrac{1}{2})\tau - T\big)$\;
                $W_l = \exp\{- i \alpha f_l (A_k \otimes \sigma^{+} + A_k^{\dagger} \otimes \sigma^{-})\,\tau/2\}$\;
                $U_{\mathrm{tot}}^{(j,k)} \leftarrow W_l\, U_S\, U_B^{(j)}\, W_l\, U_{\mathrm{tot}}^{(j,k)}$\;
            }
            Append $U_{\mathrm{tot}}^{(j,k)}$ to \textit{UTotArray}\;
        }
    }
    \KwRet \textit{UTotArray}\;
}

\Fn{\FMain}{
    Determine system unitary, $U_S = \exp\{-i H_S \tau\}$\;
    Determine Trotter step number, $M = \lceil 2T/\tau \rceil$\;
    \textit{UBArray} $\leftarrow$ \FAncilla{$\omega_{\mathrm{max}}$, $N_{\omega}$, $\tau$}\;
    \textit{UTotArray} $\leftarrow$ \FTot{\textit{UBArray}, $U_S$, $\mathcal{A}$, $N_{\omega}$, $f(t)$, $\tau$, $T$, $M$, $\alpha$}\;

    Initialize system state $\rho_S(1) = \rho_0$\;
    Initialize ancilla in $\rho_B = \ket{0}\bra{0}$\;
    Initialize logits $\boldsymbol{\theta}$ as zero array of size $(N_{\omega}, N_A)$\;
    Initialize baseline $\bm{b}$ as zero array of size $(N_{\omega}, N_A)$\;
    Initialize reward mean $\bar r = 0$, reward variance $v = 1$\;
    $E(1) = \tr{H_S\,\rho_S(1)}$\;

    \For{$n = 1$ \KwTo $n_{\mathrm{tot}}$} {
        \tcp{softmax policy over $(\omega,A)$ actions}
        $\bm{w} = \exp\{(\boldsymbol{\theta} - \max\boldsymbol{\theta})/\Theta_l\}$\;
        $\bm{p} = \bm{w}/\textstyle\sum \bm{w}$\;
        Sample action indices $(j, k)$ according to $\bm{p}$\;
        $U = \textit{UTotArray}[j, k]$\;

        \tcp{dissipative state update}
        $\rho_S(n+1) = \mathrm{tr}_B\{U(\rho_B \otimes \rho_S(n))\,U^{\dagger}\}$\;
        $E(n+1) = \tr{H_S\,\rho_S(n+1)}$\;

        \tcp{reward: per-step energy decrease}
        $r = E(n) - E(n+1)$\;

        \tcp{reward normalization with decay $\delta$}
        $\bar r \leftarrow \delta\,\bar r + (1-\delta)\,r$\;
        $v \leftarrow \delta\,v + (1-\delta)(r - \bar r)^2$\;
        $\tilde r = (r - \bar r)/(\sqrt{v} + \epsilon)$\;

        \tcp{two-timescale logit dynamics}
        $\boldsymbol{\theta}[j,k] \leftarrow \boldsymbol{\theta}[j,k] + \eta\,\tilde r$\;
        $\boldsymbol{\theta} \leftarrow \boldsymbol{\theta} + \gamma_{\mathrm{fast}}\,(\bm{b} - \boldsymbol{\theta})$\;
        $\bm{b} \leftarrow \gamma_{\mathrm{slow}}\,\bm{b} + (1-\gamma_{\mathrm{slow}})\,\boldsymbol{\theta}$\;
    }
}
\label{LearningAlgo}
\end{algorithm}

\section{Learning Rate}
\label{sec:AppendixB}

\begin{figure}
	\subfigure[$N = 4$]{
            \includegraphics[width=.3\textwidth]{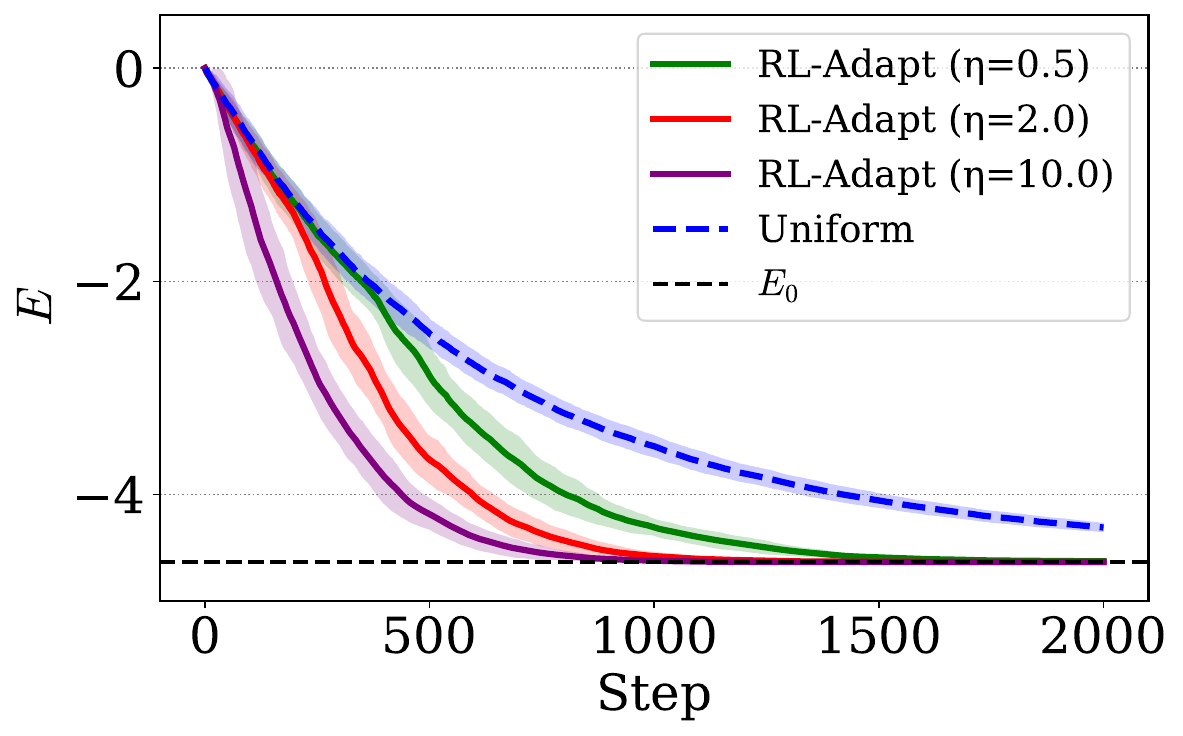}
	}
	\subfigure[$N = 6$]{
		\includegraphics[width=.3\textwidth]{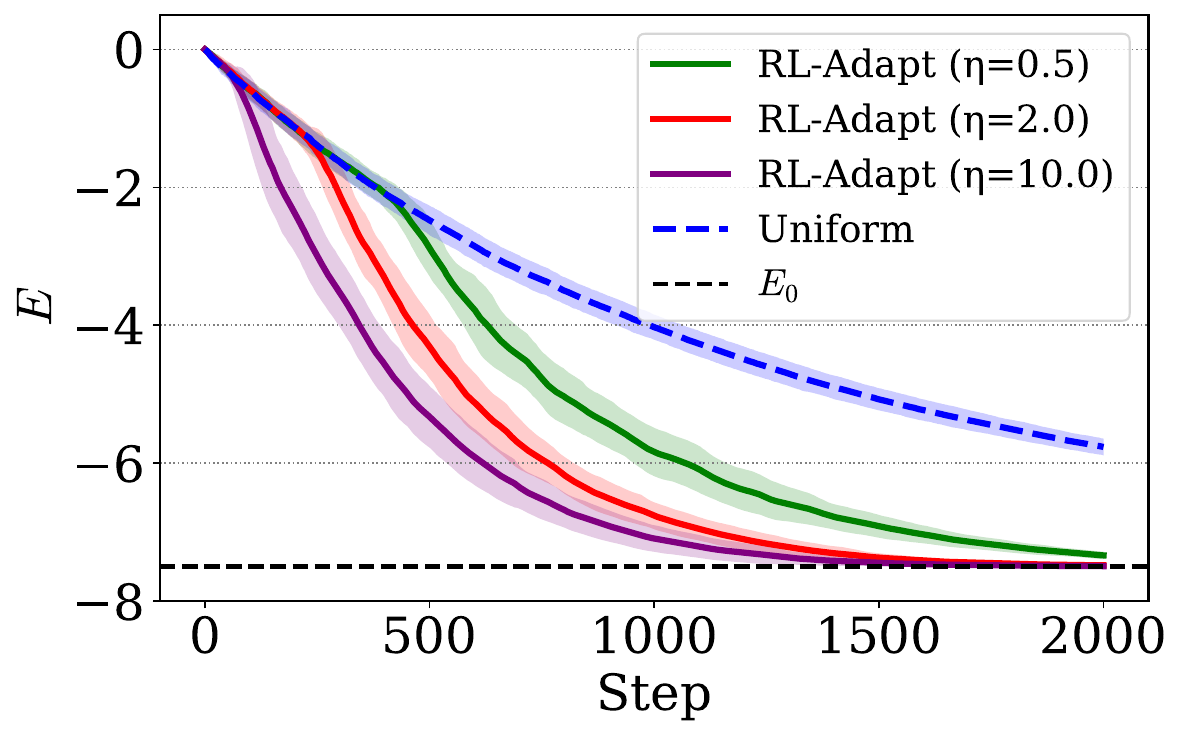}
	}
    \subfigure[$\mathrm{H}_4$]{
		\includegraphics[width=.3\textwidth]{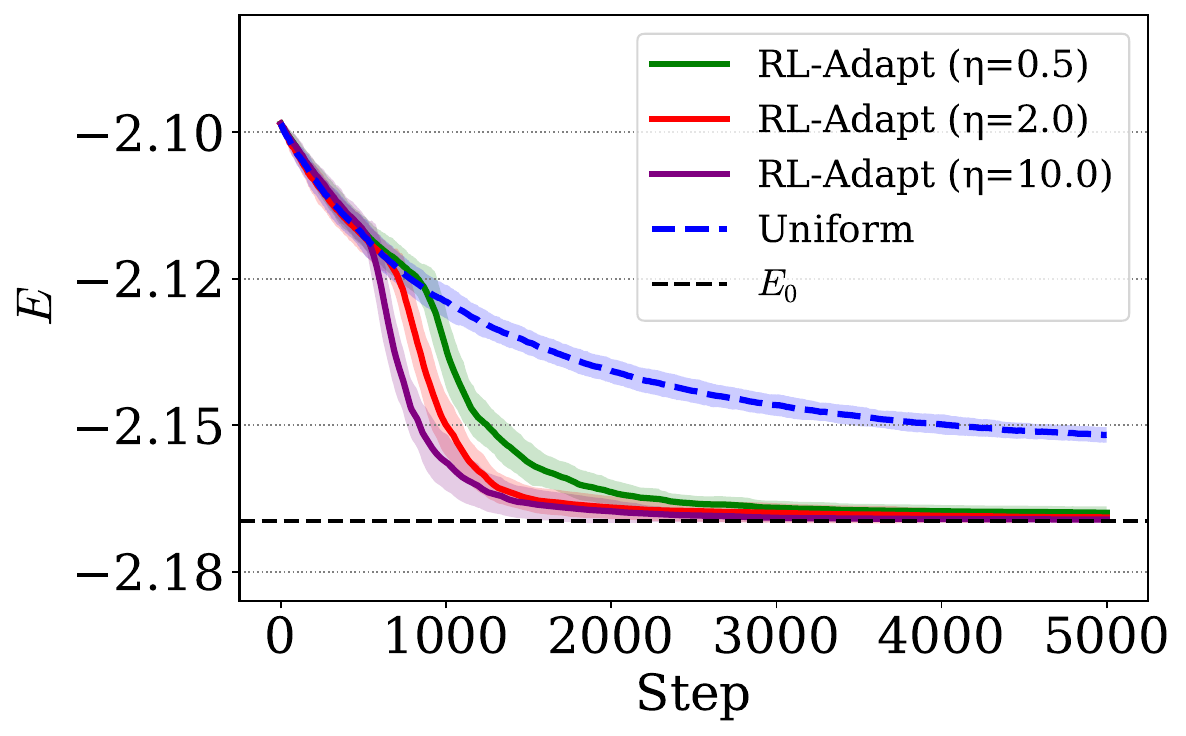}
	}
	\caption{\label{fig:EtaSweep}Energy dissipation at different learning rates, $\eta$ for the 1D XY model with (a) 4 sites and (b) 6 sites and (c) $\mathrm{H}_4$. Plotted curves are averages over 50 realizations of the algorithm, with the shaded regions corresponding to one standard deviation. Hamiltonian parameters for the XY model are $J = 1$, $h = -0.7$. RL parameters are $n_{\omega} = 40$ (XY model) and $n_{\omega} = 80$ ($\mathrm{H}_4$), $\gamma_{\mathrm{fast}} = 40/n_{\mathrm{tot}}$ and $\gamma_{\mathrm{slow}} = 2/n_{\mathrm{tot}}$.}
\end{figure}

In this appendix we examine the impacts of varying the learning rate $\eta$ on the performance of the algorithm. The learning rate controls the magnitude of the logit update, and thus how strongly the RL policy biases future action selection based on the standardized reward for the current action. In Fig.~\ref{fig:EtaSweep} we see that increasing the learning rate shortens the initial period where the RL-Adapt and non-adaptive approaches perform similarly, before the RL-Adapt method begins to display a more rapid decrease in the energy. This indicates that this initial period of similar performance between the adaptive and non-adaptive implementations is a result of the RL scheme requiring enough iterations to sufficiently bias the distribution towards the more optimal $(A, \omega)$ action pairs. Increasing the learning rate shortens this timescale by more aggressively biasing the distribution. However, for larger operator pools and frequency ranges, there is a risk that a very large learning rate counteracts the benefits of the fast residual in the two-timescale learning rule by too heavily favoring actions that give a good initial energy decrease, even once those actions are no longer beneficial since the population of the corresponding eigenstates has been exhausted.               

\section{Gap Correlation Diagnostic}
\label{sec:AppendixC}

In this appendix we provide details of how the live and dead gap correlations plotted in Fig.~\ref{fig:XYDiag} are calculated. Consider a discretized grid of $\omega$ values, $\{\omega_m\}_{m=1}^{N_\omega}$ We begin by isolating the contribution of the fast timescale learning to the logits by subtracting off the baseline contribution,
\begin{equation}
    \boldsymbol{\theta}^{\mathrm{fast}} = \boldsymbol{\theta} - \bm{b}
\end{equation}
We then sum the fast logits over the interaction operator pool and apply the softmax to determine the fast preference for each $\omega$ value, 
\begin{equation}
    q_m \;=\; \frac{\exp\!\big(g_m\big)}{\sum_{m'} \exp\!\big(g_{m'}\big)},
    \qquad
    g_m \;=\; \sum_{a \in \mathcal{A}} \theta^{\mathrm{fast}}_{a,m}.
\end{equation}
For the instantaneous state $\rho_n$, let
$p_{v} = \langle \psi_{v} | \rho_n | \psi_{v} \rangle$ denote the population of
the $v$-th eigenstate. The set of occupied (``live'') transitions is,
\begin{equation}
    \mathcal{G}^{\mathrm{live}}(\rho_n)
    \;=\;
    \bigl\{\, E_v - E_u \;:\; u < v,\; p_v > \varepsilon \,\bigr\},
\end{equation}
where $\varepsilon$ is a population threshold. In our diagnostics we have taken $\varepsilon = 10^{-3}$. For each $\omega$ value we use a binary indicator,
\begin{equation}
    \chi_m \;=\;
    \begin{cases}
        1, & \displaystyle\min_{\Delta \in \mathcal{G}^{\mathrm{live}}}
              \lvert \omega_m - \Delta \rvert < \delta,\\[8pt]
        0, & \text{otherwise.}
    \end{cases}
\end{equation}
Here $\delta$ is a tolerance for gap-frequency alignment due to the $\omega$-discretization, which we have taken to be 1.5-times the grid spacing, $\delta = 1.5 \, \omega_{\mathrm{max}}/(N_{\omega}-1)$. We then calculate the correlation at a given time step using the standard Pearson correlation coefficient between the fast preference $q$ and the indicator $\chi$ over the $\omega$ grid,
\begin{equation}
    C(\rho_n) \;=\;
    \frac{\displaystyle\sum_{m}
          \bigl(q_m - \bar q\bigr)\bigl(\chi_m - \bar\chi\bigr)}
         {\sqrt{\displaystyle\sum_{m}\bigl(q_m - \bar q\bigr)^2}\;
          \sqrt{\displaystyle\sum_{m}\bigl(\chi_m - \bar\chi\bigr)^2}}\;
\end{equation}
where $\bar q = \tfrac{1}{N_\omega}\sum_m q_m$ and $\bar\chi = \tfrac{1}{N_\omega}\sum_m \chi_m$. Lastly, we average the correlation over each independent run to verify that any positive correlation observed is not the result of a particularly successful individual run, 
\begin{equation}
    \overline{C}(n)
    \;=\;
    \frac{1}{N_{\mathrm{runs}}}
    \sum_{r=1}^{N_{\mathrm{runs}}} C^{(r)}(\rho_n),
\end{equation}

The dead-gap correlation is constructed analogously, but using $\mathcal{G}^{\mathrm{dead}}$, where $\mathcal{G}^{\mathrm{dead}}$ consists of the complementary set of all transitions not included in $\mathcal{G}^{\mathrm{live}}$.

\section{Enrichment Diagnostic}
\label{sec:AppendixD}

In this appendix we provide details of how the enrichment, as shown in Fig.~\ref{fig:XYEnrichment} is calculated. We begin by computing the matrix elements of each interaction operator in $\mathcal{A}$ in the energy eigenbasis,
\begin{equation}
    \bigl(A_j\bigr)_{vu}
    \;=\;
    \langle \psi_v | A_j | \psi_u \rangle,
\end{equation}
and define the maximal amplitude
$\mu_j = \max_{v,u} \lvert (A_j)_{vu} \rvert$. An eigenstate pair $(v,u)$ with $u < v$ is considered driven
by $A_j$ if its matrix element exceeds a relative amplitude cutoff $\kappa$,
\begin{equation}
    \mathcal{T}_j
    \;=\;
    \bigl\{\, (v,u) \;:\; u < v,\;
    \lvert (A_j)_{vu} \rvert > \kappa\,\mu_j \,\bigr\},
\end{equation}
with associated transition frequencies $\Delta_{vu} = E_v - E_u$. We define a target indicator that marks the discretized frequencies, $\{\omega_m\}_{m=1}^{N_\omega}$, that are resonant with the frequency transitions driven by $A_j$ (target frequencies), to within a tolerance $\delta$,
\begin{equation}
    \Xi_{j,m}
    \;=\;
    \begin{cases}
        1, & \displaystyle\min_{(v,u)\in\mathcal{T}_j}
              \lvert \omega_m - \Delta_{vu} \rvert < \delta,\\[8pt]
        0, & \text{otherwise.}
    \end{cases}
\end{equation}
For our diagnostic, we have taken this tolerance to be the same as used in the correlation diagnostic in Appendix~\ref{sec:AppendixC}, namely 1.5-times the grid spacing, $\delta = 1.5 \, \omega_{\mathrm{max}}/(N_{\omega}-1)$. We then apply the softmax to the baseline to determine the baseline frequency distribution for each run,
\begin{equation}
    P^{(r)}_{j,m}
    \;=\;
    \frac{\exp\!\bigl(b^{(r)}_{j,m}\bigr)}
         {\sum_{m'} \exp\!\bigl(b^{(r)}_{j,m'}\bigr)},
\end{equation}
where $b^{(r)}_{j,m}$ is the converged baseline of run $r$. The probability mass that each $A_j$ places on its own target frequencies is,
\begin{equation}
    F^{(r)}_j
    \;=\;
    \sum_{m=1}^{N_\omega} P^{(r)}_{j,m}\,\Xi_{j,m},
\end{equation}
which is compared against the uniform reference,
\begin{equation}
    F^{\mathrm{chance}}_j
    \;=\;
    \frac{1}{N_\omega}\sum_{m=1}^{N_\omega} \Xi_{j,m}.
\end{equation}
The enrichment of operator $A_j$ is then the ratio,
\begin{equation}
    \mathcal{E}^{(r)}_j
    \;=\;
    \frac{F^{(r)}_j}{F^{\mathrm{chance}}_j}.
\end{equation}
Note that $\mathcal{E}_j > 1$ indicates the slow baseline preferentially weights the frequencies actually driven by $A_j$, while $\mathcal{E}_j = 1$ corresponds to a weighting equivalent to uniform random sampling. The enrichment is then averaged over each independent run to produce the main plot in Fig.~\ref{fig:XYEnrichment},
\begin{equation}
    \overline{\mathcal{E}}_j
    \;=\;
    \frac{1}{N_{\mathrm{runs}}}
    \sum_{r=1}^{N_{\mathrm{runs}}} \mathcal{E}^{(r)}_j.
\end{equation}
For the normalized enrichment in the inset of Fig.~\ref{fig:XYEnrichment}, we begin by noting that, since $F^{(r)}_j \le 1$, the enrichment $\mathcal{E}^{(r)}_j$ is bounded
above by the value obtained when the entire baseline mass is placed on
the target frequencies. This ceiling is set by,
\begin{equation}
    \mathcal{E}^{\mathrm{max}}_j
    \;=\;
    \frac{1}{F^{\mathrm{chance}}_j}
    \;=\;
    \Biggl(\frac{1}{N_\omega}\sum_{m=1}^{N_\omega}\Xi_{j,m}\Biggr)^{\!-1},
\end{equation}
i.e. the reciprocal of the fraction of grid frequencies that are targets of $A_j$. To compare operators with differing numbers of target frequencies on a common scale, we normalize the enrichment by its ceiling,
\begin{equation}
    f^{(r)}_{\mathrm{max},\,j}
    \;=\;
    \frac{\mathcal{E}^{(r)}_j}{\mathcal{E}^{\mathrm{max}}_j}
\end{equation}
so that $f_{\mathrm{max},\,j}\in[0,1]$, with $f_{\mathrm{max},\,j}=1$
indicating that the slow baseline concentrates all of its weight on the
frequencies driven by $A_j$. As before, in the inset plot we show the run-averaged quantity,
\begin{equation}
    \overline{f}_{\mathrm{max},\,j}
    \;=\;
    \frac{1}{N_{\mathrm{runs}}}
    \sum_{r=1}^{N_{\mathrm{runs}}} f^{(r)}_{\mathrm{max},\,j},
\end{equation}

\twocolumngrid

\bibliography{AdaptDissBib}
		
\end{document}